\documentclass[useAMS,usenatbib]{mn2e}
\usepackage{graphicx}

\title[The environment of low redshift quasar pairs]{The environment of low redshift quasar pairs.}

\author[A. Sandrinelli, R. Falomo, A. Treves, E. P. Farina, M. Uslenghi] {A. Sandrinelli$^{1,2,3}
$\thanks{E-mail:asandrinelli@yahoo.it / angela.sandrinelli@brera.inaf.it}, R. Falomo$^{4}$, A. Treves$^{1,2,3}$, E. P. Farina$^{5}$, M. Uslenghi$^{6}$\\
$^{1}$Universit\`a degli Studi del$\l'$Insubria, Via Valleggio 11, I-22100 Como, Italy\\
$^{2}$INAF - Osservatorio Astronomico di Brera, Via Emilio Bianchi 46, I-23807 Merate, Italy\\
$^{3}$INFN - Istituto Nazionale di Fisica Nucleare, Sezione Milano Bicocca, \\
Dipartimento di Fisica G. Occhialini, Piazza della Scienza 3, I-20126 Milano, Italy \\
$^{4}$INAF - Osservatorio Astronomico di Padova, Vicolo dellÕ Osservatorio 5, I-35122 Padova, Italy\\
$^{5}$Max Planck Institut f\"ur Astronomie,
K{\"o}nigstuhl 17, 69117 Heidelberg, Germany\\
$^{6}$INAF-IASF - via E. Bassini 15, I-20133 Milano, Italy}

\begin{document}

\date{Accepted -- Received --}

\pagerange{000--000} \pubyear{0000}

\maketitle

\label{firstpage}

\begin{abstract}
We investigate the properties of the galaxy environment of 
a sample of 14  low redshift (z $<$ 0.85) quasar  physical pairs extracted from SDSS DR10 archives.
The pairs have a systemic radial velocity difference $\Delta V_\parallel \leqslant$ 600 $km \ s^{-1}$ 
(based on [OIII]5007 \AA \  line) and   projected distance $ R_\bot \leqslant$ 600 kpc. 
The physical association of the pairs is statistically confirmed at a level of $\sim$ 90\%.
For most of the images of these quasars we are able to resolve their host galaxies that turn out to be 
on average similar to those of quasars not in pairs.
We also found that quasars  in a  pair are on average in region of modest galaxy overdensity extending 
up 0.5 Mpc from the QSO.
This galaxy overdensity is indistinguishable from that of a homogeneous sample of isolated quasars  
at the same redshift and with similar host galaxy luminosity.
These results, albeit derived from a small (but homogeneous) sample of objects,  suggest that  the rare 
activation of two quasars with small physical separation does not require any  extraordinary environment. 
\end{abstract}


\section{Introduction}

It is widely accepted that all massive galaxies contain a supermassive black hole in their centers.
 However, only a small fraction of them become active and  for a very short time with respect to the  
 evolution time of the galaxies.  
 The   mechanism that activates and fuels the nuclei of galaxies is still not well understood. 
 The leading processes thought to be responsible for transforming a dormant massive black hole 
 into a luminous quasar (QSO) are dissipative tidal interactions and galaxy mergers 
 \citep[e.g.][and references therein]{Dimatteo2005,Callegari2011}. 
 Galaxy formation is known to be heavily influenced by the environment, with galaxies in clusters
  tending to be elliptical and deprived of most of their gaseous content \citep[e.g.][]{Silk1993,Kormendy2009}, 
  and in a number of cases  showing signature of interactions and mergers \citep[e.g.][]{Bennert2008,Mcintosh2008}.  
 What is the effective role of these interactions and of the large scale environment for the triggering 
 and/or fuelling the nuclear activity is not yet clear. 
  
 The investigation of QSO environments at various
  scales (from few kpc to Mpc) and at different cosmic epochs compared with that of normal galaxies
   still represents an important opportunity to unveil the link between nuclear activity and the immediate environment.
 The studies of galaxy clustering around quasars  \citep{Stockton1978,Yee1984,Yee1987}   
 and other active galaxies \citep[e.g.][]{Wake2004} aim to characterize the properties of the environment 
 and to compare them with the environment of non active galaxies. The emerging picture is not homogeneous. 
 Most of the papers conclude that quasars reside in regions of galaxy density that are higher than average, 
 albeit with significant difference among various objects. 
 Only in rare cases quasars are found in relatively rich environments  \citep{Stockton1978,Yee1984} 
  and  the typical environment is a modest group or a poor cluster of galaxies. 
  
  Contrasting results emerge
   when the quasar environments are compared with those of non active galaxies,
   depending on the properties  of the sample (nuclear luminosity, redshift, radio loudness, ect.).
Some  differentiation emerges from the comparison of radio loud and radio quiet quasars. 
 \citet{Ellingson1991} studied a sample of  radio loud quasars (RLQs) and  radio quiet quasars (RQQs) 
 at $0.3 < z < 0.6$ and found that the environments around RLQs are significantly denser than those around RQQs. 
 However, \cite{Fisher1996} and \cite{Mclure2001}  find no difference between the environments of RLQs and RQQs.
 More recently from the analysis of a large QSO dataset  at $z < 0.4$  from Sloan Digital Sky Survey (SDSS) \citet{Serber2006} 
  found that the overdensity around the quasars is larger than that around typical L* galaxies. 
   However, a more complete  comparison of quasars and inactive galaxies environments by 
  \cite{Karhunen2014}, that includes a matching of the samples in terms of both redshift and galaxy luminosity, 
  shows that the galaxy number density of the quasar environments is comparable to that of the inactive galaxies. 
  
Another important issue about the environment of quasars is that they  are found clustered similarly to galaxies in 
the local Universe  \citep[e.g.][and reference therein]{Croom2005,Porciani2004}. 
  At high redshift the clustering is more difficult to measure but there are indications that its strength would be
  higher than at the present epoch \citep[e.g.][]{Porciani2004}. 
The clearest sign of QSO clustering is the finding of quasar pairs,  see e.g.
the pioneering work of Djorgovski 1991 and
 the analysis on the 
 \cite{Veron2000} catalogue by \cite{Zhdanov2001}. 
  \cite {Hennawi2006} found a large sample of QSO pair  candidates in a wide range of redshift but a detailed study 
  of the environment was not carried out. 
  Most of these QSO pairs  ($\sim$ 80 \%) are at z $>$ 1 hindering  observations of their environment.
  A number of high redshift QSO pairs have  also been discovered  by \cite{Hennawi2010}. 
  
 Detailed study of the environment of a QSO pair at z $\sim$ 1.3 has been reported by \cite{Djorgovski1987}.  
  \cite{Boris2007}   investigated the environment of 4 QSO pairs at z  $\sim$ 1  with separations $\ga$ 1 Mpc.
 They found one pair  in a particularly  high-density region, some  evidences for  galaxy cluster  in the proximity
   of  other two, while one pair  appears isolated. A more systematic study was presented by 
   \cite{Farina2011} (hereafter F11) for  six low redshift physical quasar pairs from the SDSS dataset. 
   They reported  a pair in a moderately rich group of galaxies together with dynamical evidence of additional mass 
   to make the pairs bound systems. 
More recently \cite{Green2011} searched for signatures of  galaxy clusters and hot inter cluster medium associated with   
7 close ($R_\bot<$ 25 kpc) quasar pairs. Because of low quality images  they fail to 
 resolve the host galaxies and to set stringent limits to the galaxy environments. Nevertheless from their observations 
 there is no evidence that these pairs are in rich cluster environments.

Rare examples of quasar associations with more than two objects have been reported \citep{Djorgovski2007,Farina2013} 
but the limited number prevents to perform a statistical analysis.

In this paper  we explore the  galaxy environments and the dynamical properties around 14  low redshift quasar physical-pairs 
   derived from SDSS Data Release 10 (DR10)  spectroscopic and imaging datasets  . 
For these systems we perform a detailed analysis of their host galaxies and of the clustering of galaxies around the  pairs.
We are then able to compare the properties of these environments with those of an homogeneous sample of quasars 
not in pair spanning the same range of redshift and host galaxy luminosities. 
Finally from the difference of systemic velocity of each pair (derived from  [OIII]$\lambda5007$  emission lines, hereafter [OIII]) 
we set constraints on the total minimum mass of the systems based on the dynamic of the pair.
 
In this work we assume a concordant  cosmology:  
$H_0=70$ $km$  $s^{-1}$     $Mpc^{-1}$, $\Omega_m =0.30$ and $\Omega_\Lambda=0.70$.\\

  \section{The sample of quasar physical pairs}
    
We searched for quasar pair candidates from  a dataset of $\sim$ 260,000 quasars drawn from the  SDSS 
combining the quasar spectroscopic  catalogues of   \cite{Schneider2010}   and  of  \cite{Paris2013}. 
 We restricted the search to the $\sim$ 40,000  quasars with z$<$0.85, in order to derive  redshifts  from [OIII]   narrow 
 emission  line, which is  a much better indicator of the systemic  velocity of the quasar host galaxy 
  \citep{Hewett2010,Liu2013}.
  
 To search for QSO pair candidates we  computed the number  $N_{obs}$ of quasars  in the catalogue that have 
 $\Delta V_\parallel < \Delta V_{\parallel,limit}$ and  $R_\bot < R_{\bot,limit}$, where  $ \Delta V_{\parallel,limit}$ 
 and  $R_{\bot,limit}$ are fixed values, and compared with the number $N_{exp}$ of  expected random 
 association using the redshift permutation method \citep[e.g.][]{Zhdanov2001}.
 It consists in maintaining fixed the positions of  the quasars, permuting randomly the redshifts.  
 10,000 runs were performed. 
 We repeated the search with various value of $ \Delta V_{\parallel,limit}$ and  $R_{\bot,limit}$  in order to optimize
 the number of candidates with respect to the number of chance associations.
 It turns out that the best choice is $R_\bot < $ 600 kpc  and $\Delta V_\parallel <$ 600 km  $s^{-1}$. 
 For this  combination we find 26 QSO pair candidates of which only 3-4 ($\sim$14\%) are expected to be 
 false pairs (random associations). 
  
At this  stage of the selection  $\Delta V_\parallel$   was determined from SDSS redshifts. 
We  inspected  the spectra of all candidates 
to ensure that the systemic  $\Delta V_\parallel$ could be reliably derived from [OIII] lines.  
 For two dubious classifications  
 we removed two QSO pairs candidates, 
another one  for the lack of the [OIII] wavelength region in  one spectrum. 
Because of poor S/N,  8 pair candidates have the [OIII] line position hardly measurable for at least one quasar.
 For the remaining  pairs the [OIII] line position was measured with procedure described in F11,
where the centroid was evaluated  as the median of the barycenters of the line above  different flux thresholds
and the interquartile range  as uncertainty. 
 In  one case   $\Delta V_\parallel$ from [OIII] did not satisfy the condition $<$ 600 km/s, which instead was fulfilled 
by the SDSS redshifts, and the pair candidate  was removed.

\begin{table*}
 \begin{center}
 \caption{Properties of low redshift quasar pair sample.} 
 \label{sample}
 \begin{tabular}{@{}rlllllcrrrc}
  \hline\hline
Pair 	& A						&$z_A$ 	&$i_A$ 	& B					&$z_B$ 	&$i_B$	&$\Delta\theta$ 	&$R_\bot$	&$\Delta V_\parallel$  \\
 Nr	&					   	&		&[mag]	&					&		&[mag]	&[arcsec]			&[kpc]		&[$km \ s^{-1}$]	            \\
 (a)	&	(b)				  	&	(c)	& (d)		&(e)	     				&(f)		& (g)		&(h)				&(i)			&(j)		                      \\
  \hline

      	1	&J001103.18+005927.2 	& 0.4865  &19.75    &J001103.48+010032.6	& 0.4864       &20.62 & 652 		&390  	& 19   	$\pm$ 28\\
    	2	&J022610.98+003504.0 	& 0.4240  &19.54    &J022612.41+003402.2 	& 0.4239       &19.09	& 66.0 		&360	& 25   	$\pm$ 24\\
  	3	&J074759.02+431805.3	& 0.5011  &18.84    &J074759.65+431811.4	& 0.5017       &19.09	& 8.9 		&60   	& 123  	$\pm$ 18\\
 	4	&J081801.47+205009.9	& 0.2357  &17.45    &J081808.77+204910.1	& 0.2356       &18.81	& 118.1 		&440	& 36   	$\pm$ 16\\
  	5	&J082439.83+235720.3 	& 0.5353  &18.71    &J082440.61+235709.9	& 0.5368       &18.61	& 15.5 		&90		& 294  	$\pm$ 19\\
  	6	&J084541.18+071050.3 	& 0.5376  &18.73    &J084541.52+071152.3	& 0.5352	     &18.60	& 62.3 		&390	& 468  	$\pm$ 51\\
  	7	&J085625.63+511137.0	& 0.5420  &18.38    &J085626.71+511117.8	& 0.5432       &19.18	& 22.5 		&140	& 148  	$\pm$ 21\\
  	8	&J095137.01$-$004752.9& 0.6340 &20.23    &J095139.39$-$004828.7& 0.6369	     &20.02	& 49.8 		&350 	& 544        $\pm$ 23\\
  	9	&J113725.69+141101.3   & 0.7358  &20.03    &J113726.12+141111.4  	& 0.7372	     &20.53	&12.4		&90		& 238  	$\pm$ 28\\
	10	&J114603.49+334614.3 	& 0.7642  &20.11    &J114603.76+334551.9 	& 0.7615       &19.23	& 23.3 		&170	& 445        $\pm$ 38\\
 	 11	&J124856.55+471827.7	& 0.4386  &18.62    &J124903.33+471906.0	& 0.4386       &18.30	& 79.4 		&450	& 4    	$\pm$ 15\\
  	12	&J133046.35+373142.8	& 0.8141  &19.32    &J133048.58+373146.6  	& 0.8144	     &19.82 & 26.4 		&200	& 54  	$\pm$ 43\\
  	13	&J155330.22+223010.2	& 0.6413  &18.22    &J155330.55+223014.3  	& 0.6422       & 20.65	 & 5.9 		&40		& 175  	$\pm$ 12\\
 	14	&J222901.08+031139.8   & 0.8069  &21.69    &J222902.03+031024.7  	& 0.8087       &19.88	&76.5		&570	& 299  	$\pm$ 19\\
&&&&&&&&&&\\
\hline
 \end{tabular}
\end{center}
 {Notes: 
(a) Pair identification number. 
(b); (e)  SDSS quasar  name. 
(c); (f) Quasar redshifts derived from [OIII] line positions.
(d); (g)  i-band apparent magnitude  of the quasar  A and B, respectively
(h) Angular separation of the pair. 
(i) Proper traverse separation $R_\bot $. 
(j) Radial velocity difference.}
 \end{table*}

\begin{table}
\begin{center}
\caption{ Measurements of [O III]${\lambda5007}$ \AA \ emission lines based on the median of the baricenter of the line (see text).}
\label{OIII}
\footnotesize
\begin{tabular}{@{}rcc}
\hline\hline
Pair 			&	$\lambda_{[OIII]}$ (A) 	&$\lambda_{[OIII]}$ (B)  	\\	
Nr				&	[\AA]				&[\AA]			\\	
(a)				&(b)					&(c)				\\	
 \hline													
1		 	&7442.47  $\pm$  0.63 	& 7441.99  $\pm$	 0.26  \\	
2			&7129.84 $\pm$   0.44	& 7129.24   $\pm$	 0.32  \\	
3			&7515.70 $\pm$ 0.26	& 7518.77   $\pm$	 0.30  \\	
4			&6187.15 $\pm$ 0.24 	& 6186.41   $\pm$	 0.22  \\	
5	 	 	&7686.78 $\pm$ 0.37	& 7694.31   $\pm$	 0.26  \\	
6	 		&7698.28  $\pm$ 0.30	& 7686.28   $\pm$	 1.27 \\	
7	 		&7722.53 $\pm$ 0.47	& 7726.33  $\pm$	 0.20  \\	
8	   		&8180.92 $\pm$ 0.06	& 8195.76   $\pm$	 0.61  \\	
9	  		&8690.95 $\pm$ 0.33 	 & 8697.85   $\pm$	 0.70  \\	
10	 		&8832.83 $\pm$ 1.10     	& 8819.75 $\pm$	 0.04  \\		
11	 		&7202.80 $\pm$ 0.05   	& 7202.88   $\pm$	 0.32  \\	
12			&9082.84 $\pm$ 0.41	& 9084.48 $\pm$	 1.22 \\	
13			&8217.60 $\pm$ 0.26	& 8222.40 $\pm$	 0.07 \\	
14			&9047.04 $\pm$ 0.37	& 9056.06   $\pm$	 0.39  \\	
&&\\
\hline
\end{tabular}
\end{center}
{Notes:
(a)  Pair identification number.
(b); (c)  Emission line centers of  [O III]${\lambda5007}$ emission line for quasars A and B, respectively.
   }
 \end{table}

\begin{figure*}
\vspace{0cm}
\includegraphics[width=0.80\textwidth]{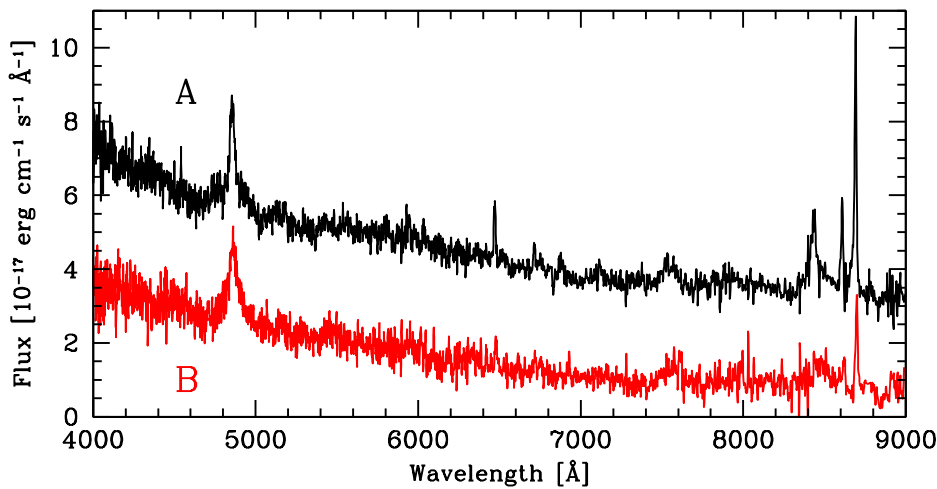}
\vspace{-6cm}
    \caption{ \label{spectra} SDSS spectra  of the QSOs in pair nr. 9 at z=0.736. 
    For clarity of comparison, the spectrum of the quasar A  is shifted 
    upwards.
    }
   \end{figure*}

In our sample of 14 QSO  pairs we then expect that 1-2 pairs could be chance superpositions. 
We can  assume that  the selected  sample is  mostly composed of physically associated objects
where the quasar velocities  are due to gravitational binding.
The final  list of the quasar pairs candidates is reported in Table \ref{sample}
and  details on [OIII] line measurements are given in Table \ref{OIII}.
None of our QSO pairs are present in catalogs of lensed quasars
(CfA-Arizona Space Telescope LEns   Survey of gravitational lenses, CASTLE\footnote{\texttt{http://www.cfa.harvard.edu/castles/}};
SDSS Quasar Lens Search, SQLS\footnote{\texttt{http://www-utap.phys.s.u-tokyo.ac.jp/$\sim$sdss/sqls/index.html}}). 
 Moreover detailed comparison  between the 
spectra of each pair exhibits clear differences that exclude the possibility of gravitational lensing.
 The redshifts of these quasar pairs  are $ 0.23\la z\la 0.82$, with  $z_{ave}=0.58 \pm 0.16$.  
An example of the SDSS spectra of a selected QSO pair is given in Figure \ref{spectra}.

\section{Host galaxies}

\begin{figure*}
  \includegraphics[width=1.2\textwidth]{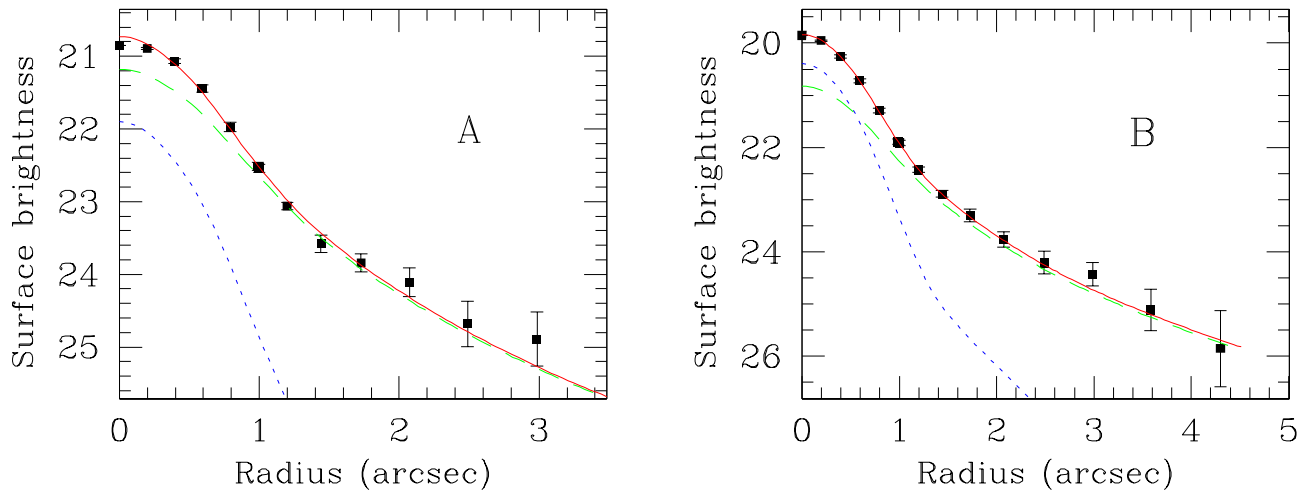}
   \vspace{-14cm}
    \caption{ \label{AIDA} Examples of  the quasar host galaxies decomposition 
   for the QSO pair nr. 2.  
   \textit{Left panel}:  Average radial brightness profile of the quasar A  (square dots)  fitted by 
    the  scaled PSF  (blue dotted  line)
   plus the the host galaxy model convolved with the PSF  (green  dashed  line), see text. 
	The  best fit is represented by the red solid line.
	 \textit{Right panel}:  The same for  the quasar B.  }
\centering
  \end{figure*}

 We retrieved the i-images of  quasar pairs from the SDSS-DR10 imaging archive. Analysis was made using the
 Astronomical Image Decomposition and Analysis software  \citep[AIDA][]{Uslenghi2008}.
 Our procedure for the study of the quasar host galaxies follows closely  that adopted by \cite{Falomo2014} 
 for the imaging study of 400 low redshift (z$<$0.5) SDSS quasars in Stripe 82. 
 In particular the analysis provides the decomposition of the quasar components, nucleus and host galaxy (see Figure \ref{AIDA}),
 resulting in  19 quasars with   resolved host galaxies (R), 5 marginally resolved (M), and 
4 objects  unresolved (U); for ten pairs we are able to characterize the host galaxy properties of either quasars.
The measured  i magnitudes (AB system) of the nucleus and of the  host galaxy  are reported in Table \ref{AIDAtable}, 
together with the rest frame  Vega $M_R$ absolute magnitude, dereddened and k-corrected. 
 Corrections for galactic extinction were taken from SDSS database
 and k-corrections from templates of  \cite{Mannucci2001} and  \cite{Francis2001} 
 for host galaxies and nuclei, respectively.
   
The average  absolute magnitudes $M_R$(nuc) of the nuclei is  -23.0 $\pm$1.1 (median -23.1 $\pm$ 0.7),
similar to the nuclear luminosities of isolated quasars  \citep[e.g.][]{Falomo2014}. 
We find that the absolute magnitude $M_R$(host) of the host galaxies range between -21 and  -24.5 
(mean -22.9$\pm$ 0.8; median -22.9$\pm$0.5).
The distribution of $M_R$(host) is comparable within  similar redshifts to that  reported for quasars   that are not in
 pairs \citep{Falomo2014}, indicating that the two families of QSO (individuals and in pairs)  
 are indistinguishable from this point of view .
Note that for $\sim$ 60\% of the objects  the observations in the i filter 
include the [OIII]  emission line. This might contaminate the measurement of the host galaxy luminosity 
(e.g. because of scattered light from the nucleus). 
However, the same effect would be present also in the case of 
single quasars  at similar redshifts.

\begin{table*}
\begin{center}
\caption{Properties of nuclei and host galaxies.}
\label{AIDAtable}
\begin{tabular}{@{}llcccccc}
\hline\hline
QSO 	 	&Class	&$i_{nuc}$	& $i_{host}$ 	&$M_R$(nuc) 	&$M_R$(host)	  	& M(host)			\\
ID		  	&		&[mag]		& [mag]	    	&	[mag]       	&[mag]		  	&[$ 10^{12} M_{\odot}$ ]	 \\
(a)		         & (b)		&(c)			&(d)			&(e)		    	&	(f)	       		 &(g)		         		         	 \\
\hline          
1A  		         &	R	&	19.90 	& 21.61   		& -22.42		&   -21.07 &	  0.09	\\	
1B  		  	&	R	&	--- 		& 20.35   		&  ---  		&   -22.33	&	  0.3		\\				
2A  			 &	R	&	21.68 	& 19.51   		& -20.38		&   -22.74	&	  0.4		\\			
2B  		 	&	R	&	20.16 	& 19.03   		&  -21.9		&   -23.22	&	  0.7		\\			
3A  			&	R	&	19.19 	& 19.55   		& -23.23		&   -23.27	&	  0.7		\\			
3B  		  	&	R	&	19.45 	& 19.90   		& -22.97		&   -22.94	&	  0.5		\\			
4A  		 	&	R	&	19.65 	& 17.92   		& -20.64		&   -22.66	&	  0.5		\\			
4B  		  	&	R	&	19.29 	& 19.68   		& -20.99		&    -20.90	&	  0.09	\\		
5A   		 	&	R	&	18.90 	& 19.76   		& -23.65		&   -23.27	&	  0.6		\\		
5B   		 	&	M 	&	18.77 	& 20.43   		& -23.79		&   -22.61	&	 0.4		\\			
6A  		 	&	R 	&	18.81	& 20.70   		& -23.74		&   -22.36	&	  0.3		\\	
6B 			&	M	&	18.72 	& 20.53   		& -23.84		&   -22.51	&	 0.3		\\	
7A  		 	&	R	&	18.68 	& 19.48   		& -23.86		&   -23.55	&	  0.9		\\
7B  		  	&	U 	&	19.17 	& ---         		& -23.37		&   ---	&	  ---	   	\\  	
8A  		 	&	R	&	20.50 	& 19.98  	 	& -22.46		&   -23.64	&	  0.9		\\		
8B  			&	R  	&	20.28 	& 21.23  	 	&  -22.7		&    -22.40 &	   0.3		\\  	
9A  		 	&	U	&	20.08 	& ---    		& -23.01		&   ---	  &	   ---		\\	
9B  			&	R	&	21.32	& 21.03    		& -22.05		&   -23.15	 &	   0.5		\\ 	
10A  		  	&	R	& 	20.34 	& 21.10   		& -23.13		&    -23.20  &	   0.5 	\\
10B  		 	&	R  	&	19.25 	& 19.69   		& -24.21		&   -24.59	 &	  1.9		\\
11A  		 	&	M	&	19.35 	& 19.34   	 	& -22.75		&   -22.98	 &	   0.5		\\
11B  		  	&	U	&	18.31	& ---         		& -23.79		&   ---    	 &	   ---	   	\\  
12A  		 	&	M 	&	19.43 	& 21.01    		 & -24.33		 &  -23.51	 &	   0.7      	\\
12B  			&	R	&	19.99 	& 20.27  	  	& -23.66		&   -24.25	 &	   1.4		\\
13A  		 	&	R	&	18.29 	& 20.04   	 	& -24.73		&   -23.65	 &	   0.9		\\
13B  		  	&	U	&	20.64 	& ---          		& -22.39		&   ---	 &	  ---  	 	\\ 
14A  		 	&	R 	&	22.05 	& 22.27   		 & -21.75		&   -22.37  &	   0.3		\\
14B  		  	&	M	&	19.99 	& 21.97    		& -23.82		&   -22.68  &	   0.3		\\
&&&&&&&\\  
\hline
 \end{tabular}
\end{center}
{ Notes. 
(a) Quasar identifier.
(b)  Resolved (R),  marginally resolved (M), unresolved (U).
(c); (d)  Apparent i-magnitude (AB system)  of the nucleus and host galaxy.
 (e); (f)  Absolute R-band magnitude (Vega system, k-corrected and dereddered) of the nucleus and host galaxy.
  (g)  Mass of the host galaxy (see text).
}
 \end{table*}

  \section{Galaxy environment of the quasar pairs }

In order to characterize the QSO pair environments, we analyzed the distribution of galaxies around the quasars using  
SDSS DR10 catalogues. From these we obtained position and photometry of galaxies  by selecting all primary 
objects photometrically classified as galaxies.
In each quasar pair field, we analyzed the i-band  surface  distribution of the galaxies within 
 a circular area  of 15 arcmin radius, corresponding to a  projected distance of 3.4  Mpc from
 the nearest pair  (z=0.236) and 9.3 Mpc from the farthest (z=0.807).

\begin{figure}
\includegraphics[width=84mm]{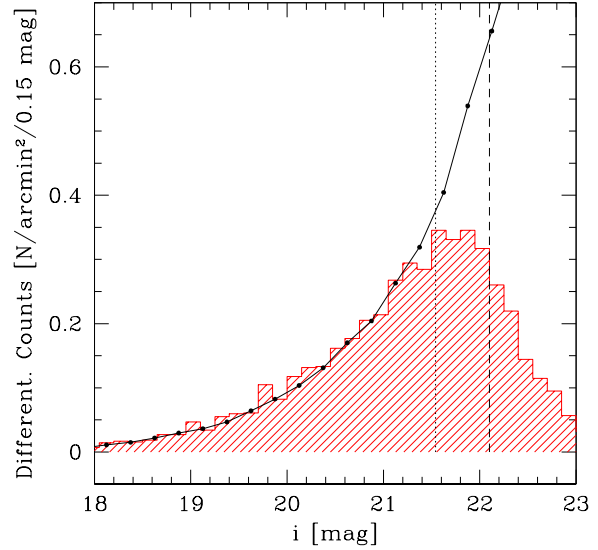}
 \vspace{-0.5cm}
  \caption{Number counts of galaxies as a function of i-magnitude in the field of QSO pair nr.2
  (red shallow histogram). The black solid line represents for comparison 
  the counts from deep survey of Capak et al. (2007).
The dotted  and dashed vertical lines mark  the  median magnitude and 50\%
completeness threshold.   
   }
     \label{distrib}
\end{figure}

 To estimate  the completeness  of the SDSS galaxy catalogues we compared  the differential number counts of  
  detected galaxies  as a function of the magnitude  with the very deep galaxy count data available from  the 
  Durham University Cosmology Group\footnote{Durham University Cosmology Group, 
 references and data in\texttt{ http://astro.dur.ac.uk/$\sim$nm/pubhtml/counts/counts.html}}. In particular
for each field we considered as  threshold   the  magnitude  $m_{i,50\%}$ where the 
completeness of observed galaxies drops to 50\% of that expected from \cite{Capak2007}
 (see Figure \ref{distrib}).

 The  apparent i-magnitude thresholds   are closely distributed   around the mean value
  of 21.96 $\pm$ 0.09 and  listed  in  Table \ref{info} with the corresponding
  absolute k-corrected magnitudes.   At these thresholds   we can observe  galaxies with magnitude 
 M*+2  at z $<$ 0.3,  M*+1   at $z\la0.5$ and  M*  at  z $\la$ 0.8.

\begin{table*}
\begin{center}
\caption{Statistics of galaxy in the QSO pair  environments.}
\label{info}
\footnotesize
\begin{tabular}{@{}lcccccc}
\hline\hline
Pair 				& $m_{i,50\%}$   &$M_{i,50\%}$	&$n_{bg}$		&$n_{bg}$		 &$G_{0,5} (A)$      		&$G_{0,5} (B) $	 \\	 
Nr				& [mag]		&[mag]			&[arcmin$^{-2}$]	&[$Mpc^{-2}$]		&					&				 \\
(a)				&(b)			&(c)				&(d)				&(e)				&(f)					&(g)				 \\
\hline
1				& 21.9 	&   -20.76 		&	3.40  $\pm$    0.10       &   26.20 $\pm$	0.80  &	1.22 $\pm$ 0.04  	& 0.68 	$\pm$ 0.02\\
2				& 22.1	&  -20.14 		&	4.01  $\pm$    0.10	   &   36.15 $\pm$	2.71  &	1.14 $\pm$ 0.09  	& 1.28 	$\pm$ 0.10\\
3				& 21.9	&  -20.86 		&	3.09  $\pm$    0.16	   &   23.16 $\pm$	1.49  &	1.04 $\pm$ 0.12  	& 1.04 	$\pm$ 0.15\\
4				& 21.9	&  -18.62 		&	2.24  $\pm$     0.29	   &   44.14 $\pm$	6.02 	&	0.89 $\pm$ 0.04  	& 1.10 	$\pm$ 0.04\\
5				& 21.9	&  -21.07 		&	3.21  $\pm$    0.09	   &   22.30 $\pm$	0.70  &	1.27 $\pm$ 0.07  	& 1.21 	$\pm$ 0.07\\
6				& 21.9	&  -21.08 		&	3.27  $\pm$    0.16	   &   22.97 $\pm$	1.39  &	0.87 $\pm$ 0.04  	& 0.92 	$\pm$ 0.07\\
7				& 21.8	&  -21.22 		&	3.15  $\pm$    0.15	   &   21.98 $\pm$	0.69  &	0.75 $\pm$ 0.15 	& 0.87 	$\pm$ 0.14\\
8				& 21.9	&  -21.67 		&	3.06  $\pm$    0.14	   &   18.40 $\pm$	0.59  &	1.11 $\pm$ 0.07	& 1.25 	$\pm$ 0.05\\
9				& 22.0	&  -22.22 		&	3.38  $\pm$    0.16	   &   17.85 $\pm$	1.05  &	1.21 $\pm$ 0.04  	& 0.93 	$\pm$ 0.05\\
10				&22.0	&  -22.41 		&	3.41  $\pm$   0.18	   &   17.40 $\pm$	1.02  &	1.27 $\pm$ 0.07  	& 1.13 	$\pm$ 0.07\\
11				& 22.0	&  -20.34 		&	3.86  $\pm$    0.10	   &   33.88 $\pm$	0.87  &	1.15 $\pm$ 0.04  	& 1.07 	$\pm$ 0.04\\
12				& 22.1	&  -22.68 		&	4.56  $\pm$    0.23	   &   22.47 $\pm$	0.98  &	0.88 $\pm$ 0.05 	& 1.51 	$\pm$ 0.06\\
13				& 22.0	&  -21.61 		&	2.89  $\pm$    0.27	   &   17.05 $\pm$	1.76  &	1.41 $\pm$ 0.02  	& 1.33 	$\pm$ 0.03\\
14				& 22.1	&  -22.63		&	4.10  $\pm$    0.27	   &   20.15 $\pm$	1.47  &	0.57 $\pm$ 0.03  	& 0.70 	$\pm$ 0.03\\
&&&&&&\\
\hline
\end{tabular}
\end{center}
{
Notes:
(a)  Quasar pair identification number .
(b)  Apparent SDSSi-magnitude threshold.
(c) Absolute magnitude  corresponding to  $m_{i,50\%}$.
(d); (e)  Background surface number density of galaxies in $arcmin^{-2}$ and in $Mpc^{-2}$, respectively. 
(f);(g) Galaxy  overdensity in the region within 500 kpc from the QSO, see text.
 } 
\end{table*}

\begin{figure*}
\centering
\includegraphics[width=0.49\textwidth]{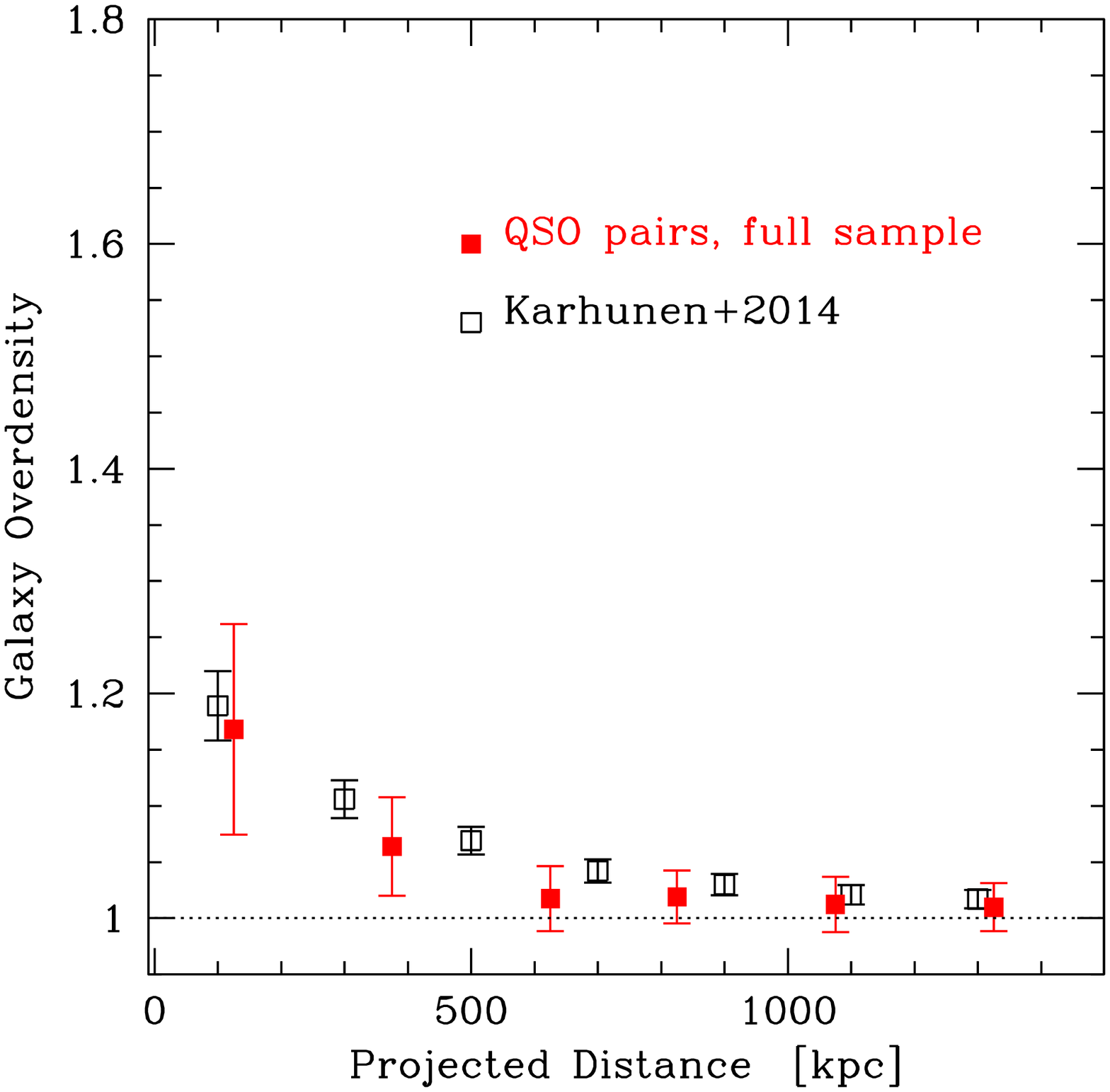}
\includegraphics[width=0.49\textwidth]{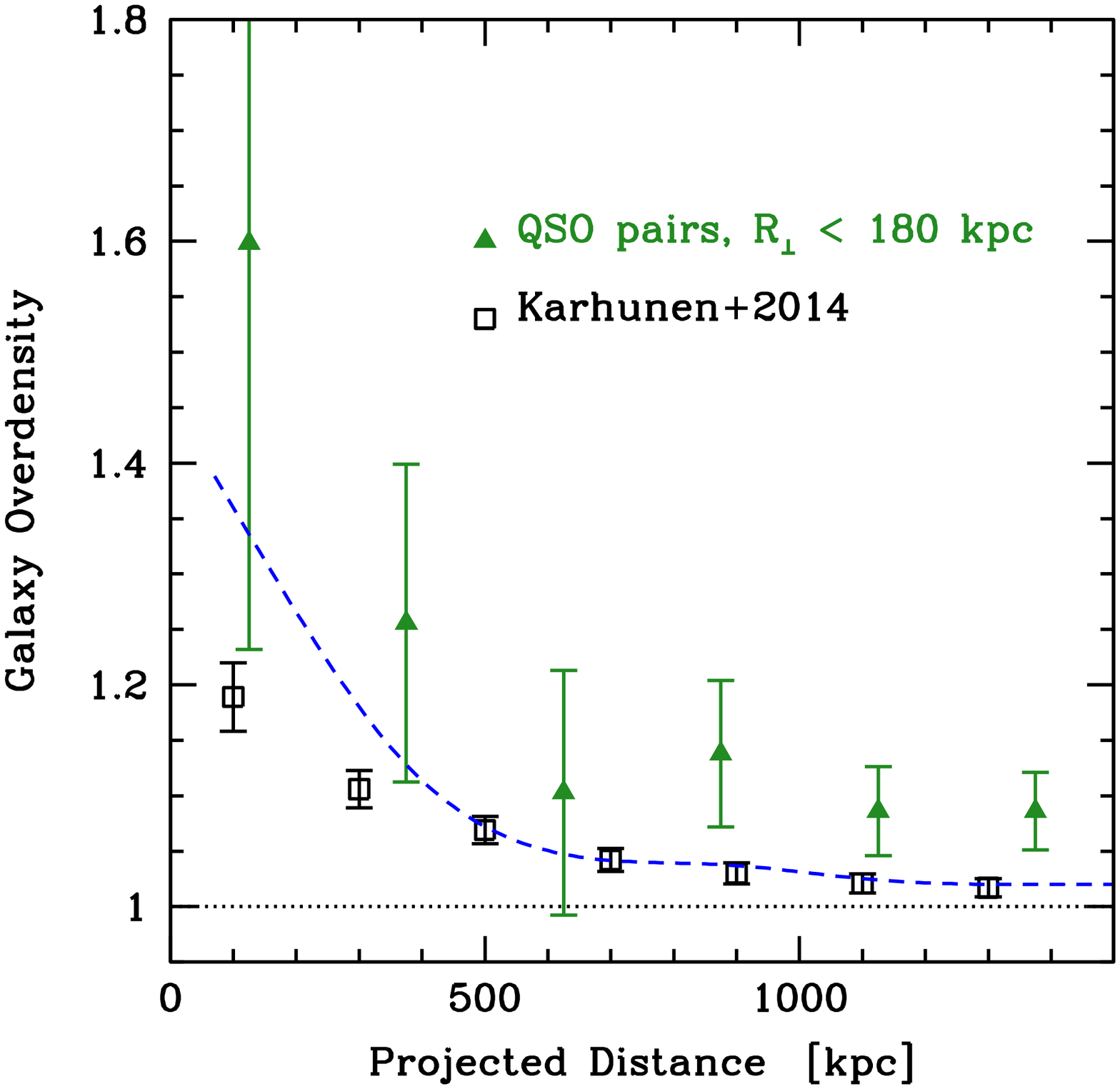}
    \caption { 
    \label{G}   \textit{Left panel:} 
    SDSS i-band mean  cumulative overdensity  of galaxies around the quasars 
    in pairs, corrected  for the superposition of the companion  environment (red filled squares)  
    as a function  of the projected distance from the quasars. 
    \textit{Right panel: } 
    Same as left panel for whole  QSO pairs with 0 $<R_\bot<$ 180  kpc (green full triangles).  
The expected galaxy overdensity around the whole QSO pairs,
derived from the galaxy overdensity in the left panel is plotted with the blue dashed line.
     In both the panels the mean cumulative overdensity around isolated quasars 
      from the subsample at i $<$ 22 mag derived by Karhunen et al. (2014) is plotted for a comparison.
     }  
\end{figure*}

In order to study the galaxy environment we followed the procedure described by \cite{Karhunen2014}.
To evaluate the  surface number  density of galaxies in the background, $n_{bg}$, we considered 
the galaxies with $i<m_{i,50\%}$ and projected angular distance between  7  and 15 arcmin
from the midpoint of the quasar pair.	
This corresponds to a  minimum distance from the QSO of $\sim$ 1.6 Mpc for the nearest target.
The region was then divided into annuli with width of 1 arcmin and
we compute  $n_{bg}$ as the median of the galaxy surface density
of  each annulus and the semi interquartile range is assumed as scatter (see Table \ref{info}).
Finally for each QSO we  counted the surface number density 
 of galaxies in a number 
of annuli with width of 250 kpc  around the target.
 The galaxy overdensity  of the QSO environment is the ratio between this number density and that of the background.
In order to take into account the contribution of  galaxies in the field around the quasar pair
 in the case that the annuli  around the two QSO  are overlapping,
   we have subdivided the excess galaxies in common to an equal number for each quasars. 
The  average cumulative overdensity distribution for the 28 quasars is reported in Figure \ref{G}, left panel,
and compared with that of  isolated quasars 
derived  by \cite{Karhunen2014}. 
 We find that on average the galaxy  overdensity around  quasars in pair is indistinguishable 
 from that of isolated quasars.
 For each QSO in our sample we report in Table \ref{info} the galaxy overdensity 
 inside a radius of 500 kpc.

It is of interest to probe whether the galaxy overdensity depends on the separation of the quasar pairs. 
To  aim this we computed the galaxy overdensity of the six pairs that are separated by less than 180 kpc 
  and compare it with that expected under the assumption that each individual QSO contributes to the average
   value of galaxies (as given in Figure \ref{G}, left panel). The comparison (see Figure \ref{G}, right panel) 
   suggests that closest quasar pairs may be in richer environments than those at larger separation. 
We performed a KS test comparing the galaxy overdensity distribution of QSO pairs with $R_{\bot} <$ 180 kpc
to that of QSO pairs with larger separations. For the cumulative galaxy overdensity up to 1500 kpc the KS test 
yields the probability p=0.08. This indicates that
the  suggestion should be confirmed by a significantly larger sample.

\section{Summary and conclusions}

We have investigated the properties of the environments of a sample of 14 physical  low redshift QSO pairs. 
We  found that  the quasars in pairs are on average in regions of modest galaxy overdensity
extending up to $\sim$ 0.5 Mpc.
This overdensity is indistinguishable from that of a  homogeneous sample of isolated quasars \citep{Karhunen2014}
that  is matched in redshift and host galaxy luminosity.
We note that for the closest QSO pairs  there is a suggestion of a larger overdensity 
roughly commensurated to the sum of the average individual  quasar environments.

Although the  number  of  known QSO  pairs candidates \citep[e.g.][]{Hennawi2006,Myers2008} 
is much larger than that in our sample, a direct comparison with other results is at present not possible.
For instance, the extended sample of QSO pairs of \cite{Hennawi2006} covers a wide redshift range 
(up to z=3, majority of pairs at z$>$1), for which a detailed environmental study is not available 
and would require a  major observational effort on large telescopes.
Recently  \cite{Green2011} studied the environments around 7 close QSO pairs ($R_{\bot}<$ 30 kpc) 
in a redshift range comparable with ours.
They  searched for extended X-ray emission as
evidence for a local group - or cluster - sized dark matter halo
associated with these quasar pairs, and found none.  
In potential contrast to our results, they didn't detect a significant
optical/infrared galaxy density enhancement. 
Moreover, due to the relatively bright magnitude limits of SDSS images at the redshift of these pairs,
 only the most luminous  galaxies possibly associated to  the quasars could be detected using their 
 DWCM technique  \citep[Distance and error-Weighted Color Mean,][]{Green2011}.
Within these limits, their results that QSO pairs avoid rich cluster environments
 are qualitatively in agreement  with our findings.

\begin{table}
\begin{center}
\scriptsize
\caption{Comparison of the minimum  virial  mass of the the QSO pairs   with  the total mass of their host galaxies.}
\label{MM}
\footnotesize
\begin{tabular}{lcc}
\hline\hline
Pair 			 	&$M_{vir,min}$ 	      	   &$M_{host,A} + M_{host,B}$ \\
Nr					&[$10^{12} M_{\odot}$]	&[$10^{12} M_{\odot}$]    \\
	(a)				&(b)								   & (c)      \\
 \hline		
1 	 	  	&(0.03  	$\pm$ 0.09)   	 		& 0.4  		\\  	
2		  	&(0.05  	$\pm$ 0.10)	 		& 1.1			\\	
3	 	  	&0.20	$\pm$ 0.06			& 1.2		\\	
4	  		&(0.13  	$\pm$ 0.12)			& 0.6			\\	
5	 	  	&1.9  	$\pm$ 0.2				& 1.0		\\
6 			&20	 	$\pm$ 4				&  0.6		\\	
7	 	  	&0.7  	$\pm$ 0.2				&  (1.7) 		\\ 	
8		 	&25	  	$\pm$ 2				&  1.2		\\	
9		  	&1.1  	$\pm$ 0.3				& (1.0)			\\	
10	 		&8	   	$\pm$ 1				&  2.4		\\	
11 	 	  	&(0.00 	$\pm$  0.02)			&  (1.1)   		\\ 	
12		  	&(0.14  	$\pm$ 0.22)			&  2.0 		\\ 
13		  	&0.30  	$\pm$ 0.04	 		 & (1.8) 		\\ 	
14		 	&12 	 	$\pm$ 2				& 0.6			\\	
&&\\
\hline
\end{tabular}
\end{center}
(a) Quasar pair identification number.
(b)  Minimum virial mass of the binary system. Values encompassed by bracket are not enough  constrain.
(c)  Total mass of the QSO host galaxies  in the pair.
In the cases where   only one quasar is resolved  (see Table \ref{AIDAtable};  in brackets), we consider as the total mass 
 the twice of the resolved quasar.
 \vspace{1cm}
\end{table}

Since the environment around pairs is relatively poor,  it is of interest  to compare it with the minimum mass of the binary system
(the two QSOs) assuming it is gravitationally bound. 
Following F11 we computed the minimum virial mass associated to each  pair and compared it with the total mass 
of the host galaxies (see Table \ref{AIDAtable}), evaluated following \cite{Kotilainen2009} and \cite{Decarli2010}.
 While in  most cases the $M_{vir,min}$ is less or similar that the total  mass of the host galaxies,
 in 3 cases (out of 14)  $M_{vir,min}$ exceeds the sum of the  masses of the hosts by a factor $\ga$ 10 (see Table \ref{MM}).  
This is suggestive of a huge dark matter contribution (see also F11). However, because of the exiguity of our sample,
 to reach a firm conclusion  on the environment and dynamical properties of QSO pairs,
  a detailed spectroscopic and imaging investigation of a larger and homogeneous sample 
   is required.

\section*{Acknowledgments}

We thank the referee for her/his constructive  report which enable us to improve our paper.

We acknowledge the  support of the Italian Ministry of Education  (grant PRIN-MIUR 2010-2011).

EPF acknowledges funding through the ERC grant ÔCosmic DawnÕ.

Funding for the creation and distribution of the SDSS Archive
has been provided by the Alfred P. Sloan Foundation,
 the Participating Institutions, the National Science Foundation, the U.S. Department of Energy,
  the National Aeronautics and Space Administration, the Japanese Monbukagakusho, 
  the Max Planck Society, and the Higher Education Funding Council for England. 
  The SDSS Web Site is \texttt{http://www.sdss.org/.}

\end{document}